\documentstyle[12pt,openbib,psfig]{article}
\newcommand{\f}{\frac}
\newcommand{\pr}{\prime}

\newcommand{\G}{\Gamma}

\newcommand{\D}{\Delta}
\newcommand{\ph}{\phi}

\newcommand{\La}{\Lambda}

\newcommand{\aph}{\alpha}

\newcommand{\om}{\omega}
\newcommand{\Om}{\Omega}
\newcommand{\ep}{\epsilon}

\def\be{\begin{equation}}
\def\ee{\end{equation}}

\begin{document}

\title{\bf  A model finding a new
Richardson potential with different scales for confinement and
asymptotic freedom, by fitting the properties of ${\D}^{++}$ and
${\Om}^{-}$.}

\author{ Manjari Bagchi $^{1,~3,~*,~\S}$, Mira Dey $^{1,~3,~*}$, Sukanta Daw $^{1,~3}$ and   Jishnu Dey $^{2,~3,~*,~
\dagger}$.}

\maketitle

\begin{abstract}
{ Phenomenological  Richardson potential has built in asymptotic
freedom (AF in short) and confinement, with only one parameter
$\La$ in the potential. But it is known that the scales of AF and
confinement are not the same. In the present work a relativistic
mean field calculation for baryons is tried out with two
parameters $\La$ and $\La^\prime$ for AF and confinement
respectively .

To test the two parameter potential we calculate the energies and
the magnetic moments, of  the triple u - quark system
(${\D}^{++}$) and the triple s - quark system (${\Om}^{-}$) and
found good values for $\La~=100~MeV$ and $\La^\prime~=~350~MeV$.
So we believe that the modified Richardson potential should have
AF scale $\La~=100~MeV$ and the confinement scale
$\La^\prime~=~350~MeV$.  }

\end{abstract}

\vskip .5cm

\noindent Keywords: Hartree-Fock~--Richardson Potential--~magnetic
moment~--~dense matter.

\vskip .5cm

$^1$ Dept. of Physics, Presidency College, 86/1 ,
College
Street, Kolkata 700 073, India\\ $^2$ UGC Research Professor, Dept.
of Physics, Maulana Azad College, 8  Rafi Ahmed Kidwai Road,
Kolkata 700 013, India
\\ $^3$ Visitor (2003), IUCAA, Pune, India \\  $\dagger$Project :
Changing Interface of Nuclear, Particle
and Astrophysics,  email : kamal1@vsnl.com.\\ $^*$ Work supported  in
part  by DST  grant no. SP/S2/K-03/2001, Govt. of India.
\\ $^\S$ email : mnj2003@vsnl.net.
\newpage
\section{Introduction}

~~~~'t Hooft suggested that  the  inverse of the number of colors
$N_c$ could be used as an expansion parameter in the otherwise
parameter free QCD theory \cite{th1}.  By the end of 90's,
properties of large $N_c$ baryons have been extensively studied by
algebraic methods for spin and isospin symmetry. It has become
possible to make a unified view on the various effective theories
such as the Skyrme model, the non-relativistic quark model and the
chiral bag model \cite{HT}. Witten \cite{wit1} suggested that for
large $N_c$ baryons a mean field description could be obtained
using a phenomenological interquark potential tested in the meson
sector\footnote{We agree with the referee that a relativistic
Faddeev calculation would be more appropriate for a 3 quark system
but do not attempt that. This is because our goal is to find an
{\it effective interaction} from finite system for {\it use} in
strange star calculations.}

Indeed self consistent baryon mass calculation is feasible with
success in the mean field level \cite{ddt} using the  Richardson
potential \cite{rich} as an interquark interaction.  Richardson
potential takes care of the two features of the qq force,
asymptotic freedom and confinement as given below :

\be
 V(r_{12}) ~= ~-\f{N_c^2-1}{2N_c}
  \f{6\pi}{33-2N_f}\left[{\La}^2
r_{12}-f({\La} r_{12})\right]\label{eq:richard}
 \ee

where $-\f{N_c^2-1}{2N_c}$ is the color contribution, $N_f$ is the
number of flavors, taken to be three, and
  \be f(t)~=~1~-~4~\int_1^\infty~\f{dq}q\f{exp(-qt)}{[ln(q^2-1)]^2+\pi^2}\label{eq:ft}
           \ee
 ${\La}$ is a parameter whose value was originally chosen as 400
MeV as for small $q^2$ the potential reduces to a linear
confinement, and the linear confinement string tension from
lattice calculation is about that value. However, the asymptotic
freedom part also has the same ${\La}$. In other words, both the
confinement and asymptotic freedom scales are chosen to be the
same in the potential equation (\ref{eq:richard}) even though  it
is known that the asymptotic freedom scale should be around 100
MeV.

Same potential has been included in the relativistic HF
calculation of strange quark matter to form compact stars. These
stars are more compact than the conventional neutron stars and
fit into the Bodmer-Witten hypothesis for the existence of
strange quark matter. They lead to systems more compact than those
derived assuming the bag model type of quark confinement
\cite{D98, Li99a, Li99b} and leads to various observable
phenomenon \cite{dorota, tbd, bd, aa, sinha, mpla, qnova, sharma,
mpla2, ApJL2}. But ${\La}$ parameter had to be within $ 100 - 200
$ MeV. This is consistent with asymptotic freedom scale but for
confinement scale, it is low.

Since the scales of these two phenomena, need not be the
 same as in the original potential., we incorporate them explicitly as ${\La}^{\prime}$ and
 ${\La}$:

\be
 V(r_{12}) ~= ~-\f{N_c^2-1}{2N_c}\f{6\pi}{33-2N_f}\left[{{\La}^{ \prime}}^2
r_{12}-f({\La} r_{12})\right]\label{eq:richardtwolam}
 \ee

${\La}^{\prime}$ corresponds to the confinement part and ${\La}$
corresponds to the asymptotic freedom part.

   Recently the  magnetic moment of ${\D}^{++}$ was derived from
experimental data \cite{cm} and it was shown in \cite{sidd} that
many models of hadrons cannot fit its value satisfactorily. We
re-visit the relativistic mean field calculation \cite{ddt} with
two parameter Richardson potential (\ref{eq:richardtwolam}).
Elsewhere the star calculation is repeated with the modified
potential (\cite{mnj}).

In this work an extensive search is performed to find the
magnetic moment and mass of the ${\D}^{++}$ and ${\Om}^-$. It is
found that for the confinement part a ${\La}^{\prime}$ of
 about 340-350 MeV and for asymptotic part a ${\La}$ of 100 MeV reproduces
 both the magnetic moment and the energy satisfactorily.

  This relatively small value of asymptotic freedom parameter (${\La}$) is in agreement with
Shifman $et~ al$ \cite{svz} ($70~MeV- 100~MeV $).

 Interestingly, in a recent paper, Radzhabov and Volkov
 \cite{volv} proposed that ${\La}^{\prime}$ should be
340 MeV.

 \section{Details of calculation}

 Assuming the quarks occupy the same orbital, the mean field ${\om}_{av}(r)$
is given by

\begin{equation}
\om_{av}(\vec r)=
-\f{{N_c}^2-1}{2N_c}\int{{\phi_{jm}}^\dagger(\vec r') V(\vec
r-\vec r')\phi_{jm}(\vec r')\vec {dr'}}\label{eq:wav}
\end{equation}

where $-\f{N_c^2-1}{2N_c}$ is the color contribution in the color
singlet state, the Fock term. The color factor for the Hartree
term is zero. The energy $E_{HF}$ is given by

 \be     E_{HF}~= ~N_c\left({\ep}-\f{1}2<{\ph}_{jm}|\om_{av}|{\ph}_{jm}>\right
 )\label{eq:ehf}
 \ee

 where ${\ep}$ is the single particle energy obtained by solving

 \be  [{\vec \aph}.{\vec p}+{\beta}m+\om_{av}(r)]{\ph}_{jm}(r)~=~ {\ep}{\ph}_{jm}(r)\label{eq:singengeqn}
 \ee

where ${\aph}$'s and ${\beta}$ are the usual Dirac matrices.
  For quarks in the lowest($1s_{1/2}$) orbital, one may write

\begin{equation}
\phi_{1s_{1/2}}(r)=~\left[\f{1}{4\pi}\right]^\f{1}{2}\pmatrix{iG(\vec
r)\chi_m \cr \vec \sigma .\hat{r}F(\vec r) \chi_m}
\label{eq:wavefn}
\end{equation}

where ${\chi}_m$ is the Pauli spinor and the eqn(5) yields the
system of coupled differential equations:

 \be  \f{dG}{dr}-(m-\om_{av}+{\ep})F~=~0    \label{eq:coupeq1}        \ee

 \be  \f{dF}{dr}+\left(\f{2}r\right)F+({\ep}-\om_{av}-m)G~=~0    \label{eq:coupeq2}        \ee

 The above equations are solved self-consistently by iteration
 since $\om_{av}$ depends on Dirac large and small components
 $G(r)$ and $F(r)$.

In eqns (\ref{eq:coupeq1}) and (\ref{eq:coupeq2}), the self
consistent single particle confining potential is a vector one,
leading to instability in the solution. The same problem was
encountered by Crater and Van Alstine \cite{cv} who suggested a
prescription of taking a half-vector half-scalar form for the
linear part. This choice also leads to a cancellation of
spin-orbit effects at long range. So we add the confining part of
the two-body potential equally to the energy and mass ( 1/2
vector, 1/2 scalar potential).

The vector and scaalar potentials are respectively:

\be
 V_{vec}(r_{12}) ~= ~-\f{N_c^2-1}{2N_c}\f{6\pi}{33-2N_f}\left[\f{{{\La}^{ \prime}}^2
r_{12}}{2}-f({\La} r_{12})\right]\label{eq:twolamvec}
 \ee

\be
 V_{scalar}(r_{12}) ~= ~-\f{N_c^2-1}{2N_c}\f{6\pi}{33-2N_f}\f{{{\La}^{ \prime}}^2
r_{12}}{2}\label{eq:twolamscal}
 \ee

The vector potential $V_{vec}(r_{12})$ is used in the expression
of $\om_{av}(\vec r)$ [equation \ref{eq:wav}] and the scalar
potential $V_{scalar}(r_{12})$ is added with the mass terms in the
coupled differential equations.

The magnetic moment ${\mu}$ is given by (see  \cite{bhaduri}):-

  \be \mu~=~-e_B ~ \f{2}3~\int_0^\infty~G(r)F(r)r^3dr  \label{eq:mu} \ee

where $e_B$ is the charge of the baryon.

\vskip .5cm
   It is convenient to express the Dirac components G(r) and F(r) as
   sum of oscillators :

 \be G(r)~=~\sum_n{C_nR_{n0}} \label{eq:g}\ee
 \be F(r)~=~\sum_m{D_mR_{m1}}  \label{eq:f}\ee

where C's and D's are coefficients. The expansion reduces the
differential equation to an eigenvalue problem. The solution is to
be found self consistently by diagonalizing the matrix and
putting back the coefficients till convergence is reached.  In
general $ R_{nl}(r)$ is given by :

\be
R_{nl}(r)~=~\sqrt{\f{2n!}{{\G}(n+l+\f{3}2)}}r^lexp(-\f{1}2r^2)L_n^{l+\f{1}2}(r^2)
\label{eq:rnl} \ee

where $ L_n^{l+\f{1}{2}}(r^2)$ are Associate Laguerre polynomials.

In the calculation, r is replaced by r/b in the expression of  $
L_n^{l+\f{1}{2}}(r^2)$ where b is the oscillator length which
need not to be the same for G(r) and F(r) and they are denoted by
$b$ and $b^{\pr}$ respectively.

Starting with a trial set of C's and D's , ${\om}_{av}$ is
calculated, a matrix is constructed and diagonalized. The new
coefficients $C_n$ and $D_n$ can be read off from the eigenvector
 with the corresponding eigenvalue ${\ep}$.
The wave functions are put back in the equations and the process
is continued till self consistency is reached.

But the problem is that the RHF solutions violate translational
invariance, since they are formed by single-particle wave
functions derived by an average potential ${\om}_{av}$. As a
consequence, the centre-of-mass momentum is not well-defined in
RHF solutions and this
 entails a spurious contribution from the centre-of-mass [CM] kinetic
energy to the total energy. Since the relative importance of this
effect increases as the number of particles decreases, it is
important that it should be corrected for systems formed of few
particles. This can be done by extending to the RHF equations the
 Peierls-Yoccoz procedure of nuclear physics. The spurious
contribution is denoted by $T_{CM}$ and the baryon mass [M] has to
be compared with the difference $E_{HF}$ - $T_{CM}$. Here this
spurious contribution has also been calculated and subtracted
from $E_{HF}$ to estimate the correct energy of the baryon.

    We have also estimated the values of r.m.s. radius $r_{av}$
    and checked if the wave functions are normalized or not.
The normalization factor is denoted by N.

      $r_{av}$ is given by:

  \be r_{av}~=~\sqrt{\int_0^{r_{max}}~\left(G(r)^2+F(r)^2\right)r^4dr} \label{eq:rav}\ee

   N is given by:

   \be N~=~\int_0^{r_{max}}~\left(G(r)^2+F(r)^2\right)r^2dr \label{eq:N}\ee

 where $r_{max}$ is the upper limit of integration and is taken as 5.0
 fm to make N $\sim$ 1.

\section{Results for  ${\D}^{++}$}

\begin{figure}[htbp]
\centerline{\psfig{figure=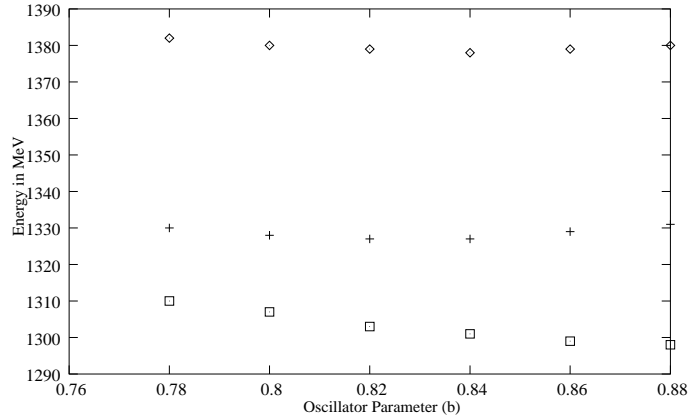,width=9cm}}
\caption{\footnotesize{Variation of Hartree-Fock energy $ E_{HF}$
with Oscillator Parameter $b$ for 5 by 5 , 7 by 7 and 9 by 9
matrices . Here $\diamond$ stands for 5 by 5 matrices, + stands
for 7 by 7 matrices, $\Box$ stands for 9 by 9 matrices . }}
\label{ffsn}
\end{figure}

We start with the simplest system, the totally symmetric spin and
isospin triple u-quark state. Here $m_u$ is taken as 4 MeV. We
checked the convergence in choosing the number of the
oscillators. In figure [\ref{ffsn}] we see that the change in
$E_{HF}$ is 32 MeV when we increase the matrix dimension from 5 X
5 to 7 X 7, but only 16 MeV from 7 X 7 to 9 X 9. It is
interesting to note that the magnetic moment remains almost same
when we increase the dimension from 7 X 7 to 9 X 9. In table
\ref{ffsn}, it is shown that $E_{HF}$ is almost independent of
oscillator parameter $b$ and $b^{\pr}$ and the mass after CM
correction is 1171 MeV.

\begin{table}[htbp]

\caption{Variation of Hartree Fock energy, centre of mass kinetic energy, mass and magnetic moment
  of ${\D}^{++}$ with oscillator parameter $b$ where
  ${\La}~={\La}^{\pr}~=350~MeV$; for 9 X 9
  matrix.}

\vskip .6cm

\begin{center}

\begin{tabular}{|c|c|c|c|c|c|c|c|}

\hline $b^{\pr}$&$b$&$E_{HF}$(MeV)&$T_{CM}$(MeV) & M(MeV)
&$\f{{\mu}}{\mu}_0$&$r_{av}$(fm)&$N$\\ \hline 0.60&0.78 &1310 &139
&1171 &6.75 &1.30 &0.988152 \\ \hline 0.60&0.80 &1307 &138 &1169
&6.69 &1.31
&0.990868 \\ \hline 0.60&0.82 &1303 &138 &1165 &6.56 &1.32 &0.995152 \\
\hline 0.60&0.84 &1301 &130 &1171 &6.42 &1.31 &0.998549 \\ \hline
0.60&0.86 &1299 &130 &1169 &6.30 &1.31 &1.000855\\ \hline
0.60&0.88 &1298 &127 &1171 &6.23 &1.30 &1.002296\\ \hline
\end{tabular}

\end{center}

\label{mat9}

\end{table}

 Now, we vary ${\La}$  and ${\La}^{\pr}$ independently. We give details of ${\La}$ and
${\La}^{\pr}$ variational results in table \ref{mtl} and table
\ref{mtlp} for matrix dimension 7 X 7 as an example. We find that
$E_{HF}$ varies widely with ${\La}^{\pr}$ as expected since
${\La}^{\pr}$ is the confining parameter. With ${\La}^{\pr} = 350
~MeV$ and ${\La} = 100~MeV$ ( table [\ref{mtl}] ) $\D $ mass =
1250 MeV and its magnetic moment 5.77 magneton. Our final result
is given in table [\ref{mm3}] for 9 X 9 matrix. We get the $\D $
mass to be 1224 MeV and its magnetic moment ${\mu}=~6.15$
magneton.

\begin{table}[htbp]

\caption{Variation of Hartree Fock energy, centre of mass kinetic energy, mass and magnetic moment
  of ${\D}^{++}$ with asymptotic parameter ${\La}$ where
  ${\La}^{\pr}$ is 350 MeV; $b^{\pr}$ is 0.60, $b$ is 0.84 using  seven by seven
  matrix. }

\vskip 0.5cm

\begin{center}

\begin{tabular}{|c|c|c|c|c|c|c|}

\hline${\La}$(MeV) &$E_{HF}$(MeV)&$T_{CM}$(MeV) & M(MeV)
&$\f{{\mu}}{\mu}_0$&$r_{av}$(fm)&$N$\\ \hline
 100&1391&141 &1250&5.77&1.11&1.000001 \\ \hline
250&1340 &141 &1199&5.95
&1.13 &1.000001 \\ \hline 300&1333 &140 &1193 &5.99 &1.14 &1.000000 \\
\hline 325&1330 &140 &1190 &6.00 &1.14 &1.000000 \\ \hline
375&1325 &140 &1185 &6.02 &1.14 &1.000000 \\ \hline

\end{tabular}

\end{center}

\label{mtl}

\end{table}

\begin{table}[htbp]

\caption{Variation of Hartree Fock energy, centre of mass kinetic
energy, mass and magnetic moment
  of ${\D}^{++}$ with confinement parameter ${\La}^{\pr}$ where
  ${\La}$ is 350 MeV; $b^{\pr}$ is 0.60, $b$ is 0.84 using seven by seven
  matrix. }

\vskip 1cm

\begin{center}

\begin{tabular}{|c|c|c|c|c|c|c|}

\hline${\La}^{\pr}$(MeV) &$E_{HF}$(MeV)&$T_{CM}$(MeV) & M(MeV)
&$\f{{\mu}}{\mu}_0$&$r_{av}$(fm)&$N$\\ \hline 250&1003 &140 &863
&6.03
&1.39 &0.999999 \\ \hline 300&1171 &139 &1032 &6.09 &1.22 &1.000001 \\
\hline 325&1251 &140 &1111 &6.03 &1.17 &1.000001 \\ \hline
375&1399 &140 &1259 &6.16
&1.14 &1.000000 \\ \hline 400&1475 &142 &1333 &6.76 &1.26 &1.000000\\
\hline

\end{tabular}

\end{center}

\label{mtlp}

\end{table}

\begin{table}[htbp]

\caption{Hartree Fock energy, centre of mass kinetic energy, mass
and magnetic moment
  of ${\D}^{++}$ where ${\La}^{\pr}$ is 350 MeV,  ${\La}$ is 100 MeV;
  $b^{\pr}$ is 0.60, $b$ is 0.84 using nine by nine
  matrix. }

\vskip .5cm

\begin{center}

\begin{tabular}{|c|c|c|c|c|c|c|}

\hline $E_{HF}$(MeV)&$T_{CM}$(MeV) & M(MeV)
&$\f{{\mu}}{\mu}_0$&$r_{av}$(fm)&$N$\\ \hline
 1354&130 &1224 &6.15&1.22&0.998476 \\ \hline

\end{tabular}

\end{center}

\label{mm3}

\end{table}

\vskip 1cm

\newpage

\begin{figure}[htbp]
\centerline{\psfig{figure=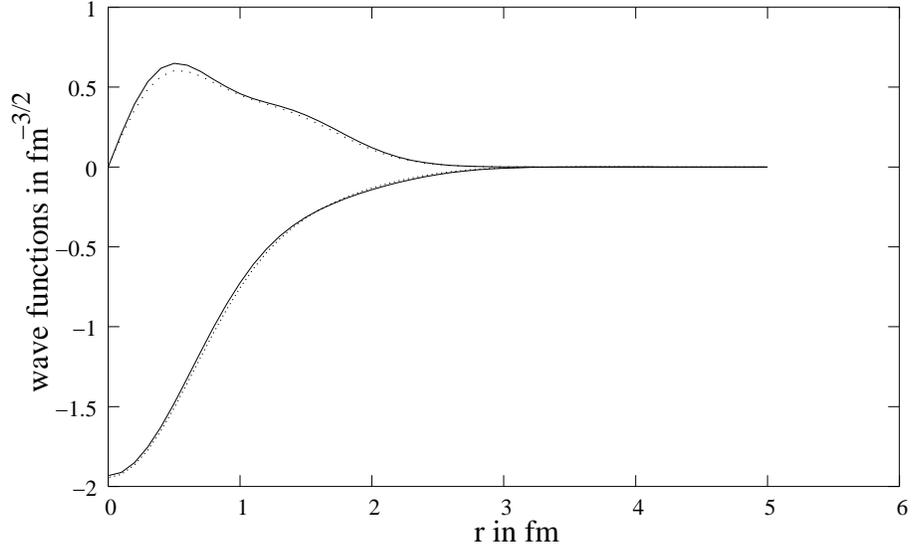,width=12cm}}
\caption{\footnotesize{Variation of the wave functions with r for
two different values of the asymptotic freedom parameter ${\La}$ (
using seven by seven matrices ); the upper ones are G(r) and the
lower ones are F(r). Here the dotted curve is for
${\La}~={\La}^{\pr}=~350~MeV $and the solid one is for
${\La}~=~100,  {\La}^{\pr}=~350~MeV$.}} \label{aswvfn}
\end{figure}

\begin{figure}[htbp]
\centerline{\psfig{figure=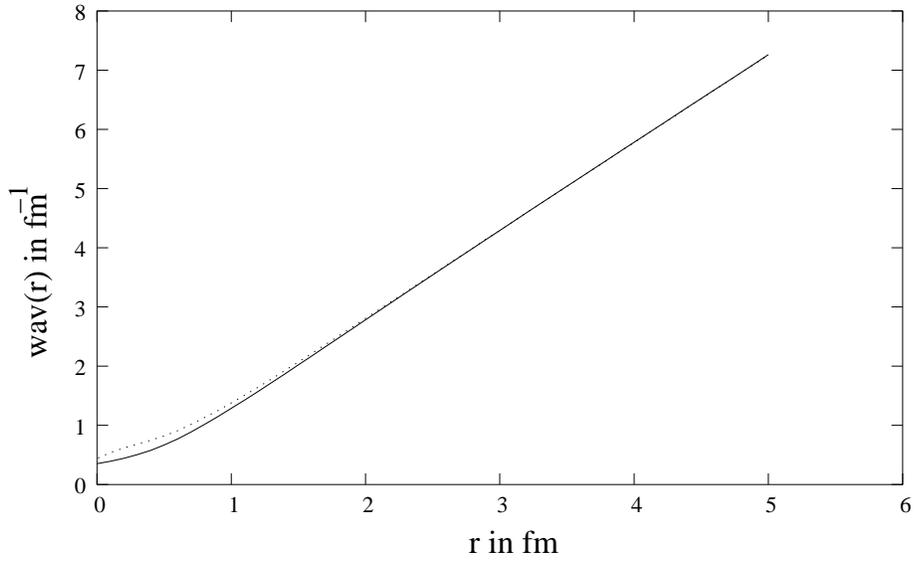,width=12cm}}
\caption{\footnotesize{Variation of the single particle potential
with r for two different values of the asymptotic freedom
parameter ${\La}$ ( using seven by seven matrices ). Here the
dotted curve is for ${\La}~={\La}^{\pr}=~350~MeV $and the solid
one is for ${\La}~=~100,  {\La}^{\pr}=~350~MeV$.}} \label{asspp}
\end{figure}

\begin{figure}[htbp]
\centerline{\psfig{figure=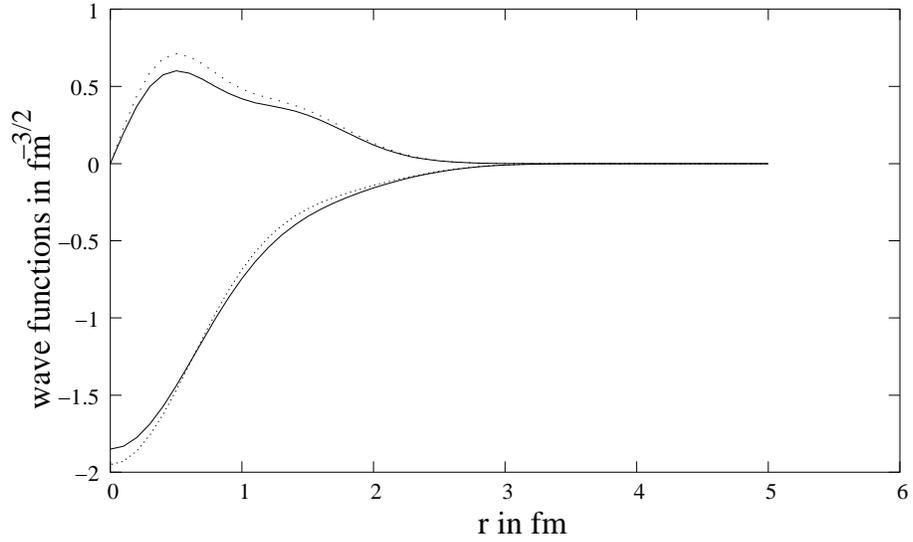,width=12cm}}
\caption{\footnotesize{Variation of the wave functions with r for
two different values of the confinement parameter ${\La}^{\pr}$ (
using seven by seven matrices ); the upper ones are G(r) and the
lower ones are F(r) . Here the dotted curve is for
${\La}^{\pr}~=~375~MeV$, ${\La}~=~350~MeV$ and the solid one is
for ${\La}^{\pr}~=~325~MeV$, ${\La}~=~350~MeV$ }.} \label{conwvf}
\end{figure}

\begin{figure}[htbp]
\centerline{\psfig{figure=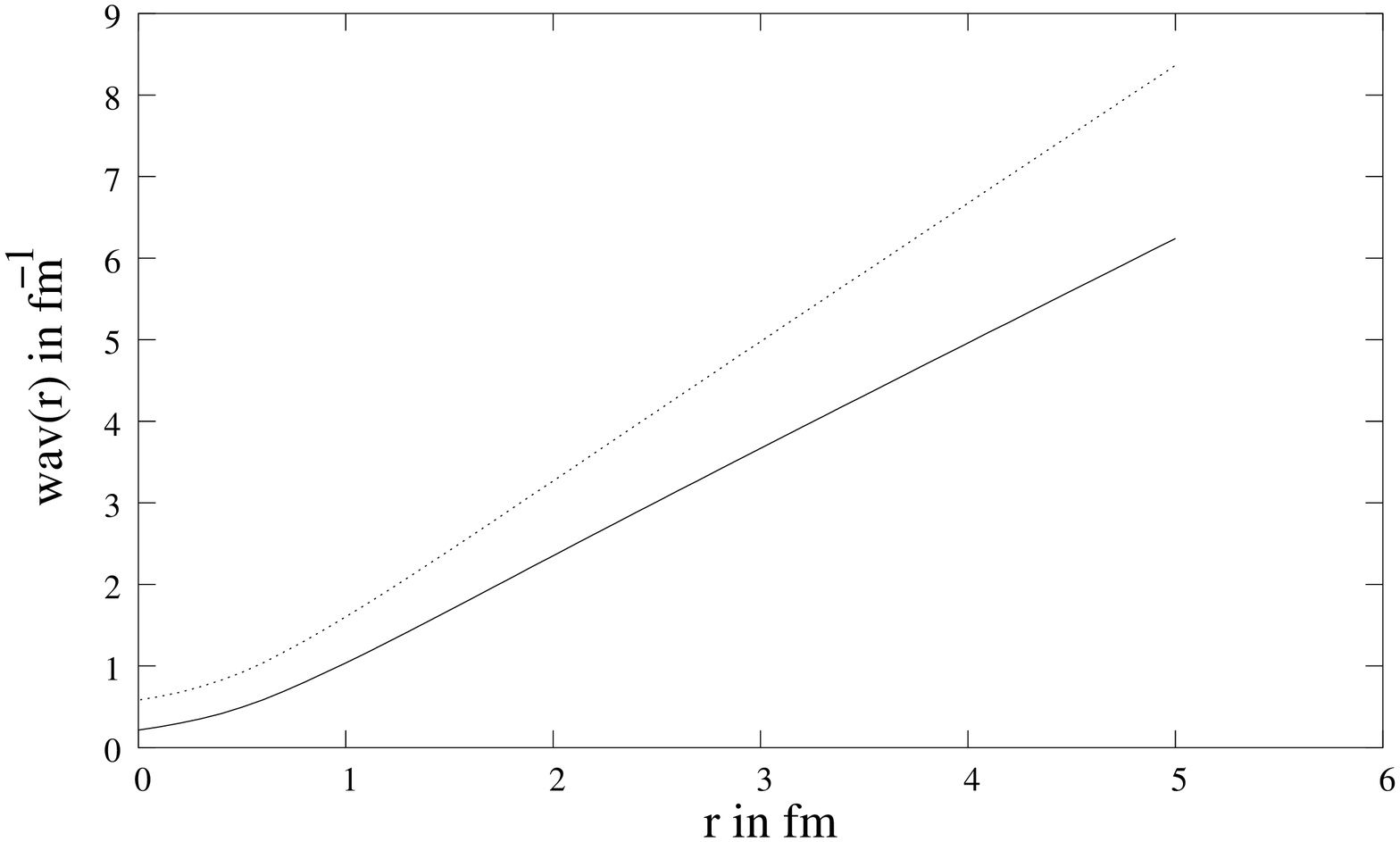,width=12cm}}
\caption{\footnotesize{Variation of the single particle potential
with r for two different values of the confinement parameter
${\La}^{\pr}$ ( using seven by seven matrices ) . Here the dotted
curve is for ${\La}^{\pr}~=~375~MeV$, ${\La}~=~350~MeV$ and the
solid one is for ${\La}^{\pr}~=~325~MeV$, ${\La}~=~350~MeV$ }.}
\label{conspp}
\end{figure}

Figures (\ref{aswvfn}) and (\ref{asspp}) show  the wave functions
and the single particle potential for ${\La}^{\pr}=350~MeV$ but
${\La}=350~MeV$ and $100~MeV$. As expected, there is no change in
large r but an insignificant variation in small r. This is
reflected in $E_{HF}$ and ${\mu}$ ( table \ref{mtl}).

On the other hand, ${\La}^{\pr}$ variation ( figures \ref{conwvf}
and \ref{conspp}) is quite significant and larger in large r.
This is reflected in $E_{HF}$ and ${\mu}$ ( table \ref{mtlp}).

 So we can conclude that for finite system calculations, the value of
the confinement parameter ${\La}^{\pr}$ plays a more important
role in the Richardson Potential.
\newpage

\section{Result with ${\Om}^{-}$}
${\Om}^{-}$ is the triple s-quark state. There is no qualitative
change in the procedure except for for the quark mass; $m_s$ is
taken here as $150~MeV$. Results for ${\La}^{\pr}={\La}=~350~MeV$
is shown in table [\ref{matn7}].

\begin{table}[htbp]

\caption{Variation of Hartree Fock energy, centre of mass kinetic
energy, mass and magnetic moment
  of ${\Om}^-$ with oscillator parameter $b$ where
  ${\La}={\La}^{\pr}=350~MeV$, $b^{\pr}$ is 0.60.}
\vskip .6cm

\begin{center}

\begin{tabular}{|c|c|c|c|c|c|c|c|}

\hline $b$&$E_{HF}$(MeV)&$T_{CM}$(MeV) &M(MeV)
&$\f{{\mu}}{\mu}_0$&$r_{av}$(fm)&$N$\\ \hline
 0.60&1639 &176&1463 &-1.95 &0.836879&1.000001 \\ \hline
 0.62 &1638 &175&1463&-1.95 &0.836778&1.000000\\\hline
 0.64 &1638 &174&1464&-1.95 &0.836879 &1.000000 \\ \hline
 0.66&1638 &172&1466&-1.95 &0.836854 &1.000001 \\ \hline
 0.68&1638 &171&1467&-1.95 &0.836882 &1.000000 \\ \hline
 0.70&1637 &169&1468&-1.95 &0.83643&0.999999 \\ \hline
 0.72&1639 &167&1472&-1.95 &0.83695 &1.00001 \\ \hline
\end{tabular}

\end{center}

\label{matn7}

\end{table}

We improve the fit by taking  different values for ${\La}^{\pr}$
and ${\La}$; specifically for ${\La}^{\pr}$ = 350 MeV and $
{\La}$ = 100 MeV.

\begin{table}[htbp]

\caption{Hartree Fock energy, centre of mass kinetic energy, mass
and magnetic moment
  of ${\Om}^{-}$ where ${\La}^{\pr}$ is 350 MeV,  ${\La}$ is 100 MeV;
  $b^{\pr}$ is 0.60, $b$ is 0.70. }

\vskip .5cm

\begin{center}

\begin{tabular}{|c|c|c|c|c|c|c|}

\hline $E_{HF}$(MeV)&$T_{CM}$(MeV) & M(MeV) &$\f{{\mu}}{\mu}_0$&$r_{av}$(fm)&$N$\\
\hline
 1721&165 &1556 &-1.92&0.849015&0.999999 \\ \hline

\end{tabular}

\end{center}

\label{om2}

\end{table}

\vskip .5cm

The centre of mass kinetic energy $T_{cm}$ is rather high
resulting a lower mass of ${\Om}^-$
\newpage

\section{Conclusions and summary}

In summary we have shown that a RHF calculation can be done for
the energies and magnetic moments of the simplest baryons
${\Om}^{-}$ and ${\D}^{++}$. We employed the oscillator basis with
good convergence. Optimized oscillator parameters are used for the
large and small components of the RHF wave functions. We belive
that this model may be good base for further investigations of
baryon properties. The center of mass correction is around 150 MeV
for these baryons (about $10$\% of $E_{HF}$). Quark masses chosen
are 4 and 150 MeV for u and s quark respectively. For q q
interaction we chose Richardson potential. We separated out the
confinement and asymptotic freedom scale parameter (${\La} ^{\pr}$
and $\La$ ). From the best fit of the energies of ${\Om}^{-}$ and
${\D}^{++}$ we find $\La ^{\pr} = 350~MeV$ and $\La = 100 ~MeV$.
We believe these are realistic values. Above parameters are now
used in strange quark matter calculation (\cite{mnj}).

\vskip .6cm
 Acknowledgments :-
\vskip .2cm
 MB, MD, SD and  JD  thank IUCAA for a short pleasant stay in
 2003. We acknowledge that the suggestion of doing a calculation
 with two ${\La}$ came originally from Dr. Ignazio Bombaci of Pisa.
 We thank Monika Sinha for careful reading of the manuscript.

%%%%%%%%%%%%%%%%%%%%%%%%%%%%%%%%%%%%%%%%%%%%%%%%%%%%%%%%%%%%%%%%%%%%%%%%%

%%%                REFERENCES

%%%%%%%%%%%%%%%%%%%%%%%%%%%%%%%%%%%%%%%%%%%%%%%%%%%%%%%%%%%%%%%%%%%%%%%%%

\end{document}